\documentclass[12pt]{article}
\usepackage{graphicx,amsfonts,amsmath,amssymb,mathrsfs,epsfig}

\title{The Analogue of Bohm--Bell Processes on a Graph}

\author{ 
Roderich Tumulka\footnote{Mathematisches Institut,
    Eberhard-Karls-Unversit\"at, Auf der Morgenstelle 10, 72076
    T\"ubingen, Germany.  E-mail:
    tumulka@everest.mathematik.uni-tuebingen.de}
}
\date{August 24, 2005}

\newcommand{\conf}{\mathcal{Q}}
\newcommand{\Q}{\conf}

\renewcommand{\Im}{\mathrm{Im}}

\newcommand{\PPP}{\mathbb{P}}
\newcommand{\RRR}{\mathbb{R}}
\newcommand{\CCC}{\mathbb{C}}

\newcommand{\ZZZ}{\mathbb{Z}}

\newcommand{\prob}{\mathrm{Prob}}
\renewcommand{\sp}[2]{\langle #1|#2 \rangle}
\newcommand{\Laplace}{\Delta} %
\newcommand{\graph}{\mathscr{G}}
\newcommand{\vertices}{\mathscr{V}}
\newcommand{\edges}{\mathscr{E}}
\newcommand{\edin}{e}
\newcommand{\aedin}{f}

\begin{document}
\maketitle
\begin{abstract}
  Bohm--Bell processes, of interest in the foundations of quantum
  field theory, form a class of Markov processes $Q_t$ generalizing in
  a natural way both Bohm's dynamical system in configuration space
  for nonrelativistic quantum mechanics and Bell's jump process for
  lattice quantum field theories.  They are such that at any time $t$
  the distribution of $Q_t$ is $|\psi_t|^2$ with $\psi$ the wave
  function of quantum theory. We extend this class here by introducing
  the analogous Markov process for quantum mechanics on a graph (also
  called a network, i.e., a space consisting of line segments glued
  together at their ends). It is a piecewise deterministic process
  whose innovations occur only when it passes through a vertex.

\medskip

\end{abstract}

\section{Introduction}

We consider quantum mechanics on a \emph{graph} $\graph$ (also called
a \emph{network}), i.e., on a topological and metric space consisting
of one-dimensional manifolds glued together at their end points \cite{kuchment}. (It
is not necessary for our purposes to regard the graph as embedded in
$\RRR^n$.) We denote by $\graph$ the set of all points belonging to
the graph: vertices and non-vertices together. The wave function at
time $t$ is a function $\psi_t: \graph \to \CCC$ on the graph (though
one could also think of functions to $\CCC^n$) and evolves according
to the usual nonrelativistic Schr\"odinger equation
\begin{equation}\label{schr}
  i \hbar \frac{\partial \psi_t}{\partial t} = - \tfrac{\hbar^2}{2}
  \Laplace \psi_t + V \psi_t \,,
\end{equation}
understood in a suitable way (see Section 2). We introduce a Markov process
$(Q_t)_{t\in\RRR}$ in $\graph$ associated with a wave function $\psi$
obeying \eqref{schr}.

This process is a contribution to the research program of providing
for every quantum theory a canonical Markov process in its
configuration space.  Examples of such processes are: the motion of
the configuration $Q_t$ in Bohmian mechanics
\cite{Bohm52,DGZ92,holland,survey} for nonrelativistic quantum
mechanics, a dynamical system (and thus a deterministic process) in
Euclidean space; Bell's process \cite{Bell86,crex1,schwer,Colinthesis}
for lattice quantum field theory, a Markovian pure jump process on a
lattice; and the Markov processes employed in ``Bell-type quantum
field theories'' \cite{crea1,crlet,crea2A,crea2B,schwer}, whose paths
are piecewise Bohmian trajectories interrupted by stochastic jumps,
and which can be regarded as a continuum analogue of Bell's process,
or as particle creation and annihilation added to Bohmian mechanics.
To all of these processes we refer as ``Bohm--Bell processes''; see
\cite{schwer} for an introduction. They are ``guided'' by the quantum
wave function $\psi$ in the sense that the transition probabilities
are determined by $\psi$, and they are \emph{equivariant processes}
\cite{crea2A,crea2B,DGZ92}, i.e., such that at any time $t$ the
distribution of $Q_t$ is $|\psi_t|^2$.

The investigation of Bell-type quantum field theories in
\cite{crea1,crlet,crea2A,crea2B,schwer,Colinthesis} has sparked
interest in natural classes of processes generalizing both Bohmian
mechanics and Bell's process. One such class, not yet exactly defined
but outlined in \cite[Sec.~5.3]{crea2B}, providing a canonical Markov
process for a given Hamiltonian, configuration space, and wave
function, is the class of ``minimal processes.'' I strongly expect
that the process on the graph we are discussing is contained in this
class, and in anticipation I call the process the ``minimal graph
process.''

What is novel about the minimal graph process is that graphs do not
belong to the spaces on which such processes have been considered
before, which are: Euclidean spaces \cite{Bohm52}, discrete
coarse-grainings of Euclidean spaces \cite{Sudbery},
infinite-dimensional vector spaces (of field variables) \cite{Bohm52},
Riemannian manifolds \cite{jamesthesis,topid1A}, lattices
\cite{Bell86}, countable unions of disjoint Euclidean spaces
\cite{crea1} or of disjoint Riemannian manifolds
\cite{crea2B,crlet,crea2A}, and manifolds with boundaries
\cite{schwer}.

The crucial difference between graphs and manifolds is that in a
vertex of a graph three or more edges can meet, forming a
\textsf{Y}-shaped (or, for more than three edges, $+$-shaped or
$*$-shaped) neighborhood of the vertex that is forbidden in a
manifold. It is exactly these \textsf{Y}-shaped neighborhoods that are
the source of a feature of the minimal graph process which is absent
for the corresponding process associated with the Schr\"odinger
equation \eqref{schr} on a manifold: while the latter is
deterministic, the minimal graph process is not, as it typically makes
a random turn at every vertex, i.e., it selects at random one among
the edges ending at this vertex, and moves away along that edge. The
random turns at the vertices constitute, in fact, the only
stochasticity in the process: once the process is on some edge, it
moves deterministically, like the Bohmian motion on a manifold, until
it arrives at a vertex, and the only random decision taken there is
along which edge to proceed. In particular, the paths of the minimal
graph process are continuous. As we show in Section 3, a deterministic
equivariant process with continuous paths is generically impossible on
a graph; thus, the \emph{topology enforces stochasticity}.  An
analogous connection between topology and stochasticity has been
observed in \cite[Sec.~6]{schwer}, in that case concerning
\emph{boundaries} of (rather than \textsf{Y}-shapes in) the
configuration space.

The scope of this paper is modest. We define the minimal graph
process, point out in what sense it is the unique analogue of
Bohm--Bell processes, and compare it to Bohm--Bell processes and some
of their limiting cases.  We regard it as a natural mathematical
extension of the class of ``minimal processes'' associated with
quantum theories, beyond the realm in which it was considered so far.
I think that the simplicity of the minimal graph process adds to the
overall picture of naturalness of the ``minimal processes'', and thus
to the confidence with which one can propose physical theories
employing minimal processes.

The paper is organized as follows. In Section~2 we define the minimal
graph process. In Section~3 we show that the minimal graph process is
uniquely selected by four postulates, and we discuss other equivariant
processes on the graph. In Section~4 we compare the minimal graph
process to the known jump processes. In Section~5 we show that the
minimal graph process is a limiting case of Bell's process. In
Section~6 we discuss how it relates to a suitable limiting case of
Bohmian mechanics. In Section~7 we consider the behavior of the
process under symmetries. In Section~8 we put the connection between
topology and stochasticity into perspective.

\section{The Minimal Graph Process}

If one tries to find an analogue of Bohmian mechanics on a graph, the
simplest process one could choose is the minimal graph process.  Let
$\vertices$ be the set of vertices of the graph $\graph$, $\edges$ the
set of edges, and $\edges_q$ for $q \in \vertices$ the set of edges
ending at $q$. We assume (without loss of generality) that an edge
cannot end at the same vertex on both sides; i.e., every closed path
consists of at least two edges. Every edge is isometric to either an
interval $[0,a]$ of positive length or the half-infinite interval
$[0,\infty)$. We assume that the graph is connected and has only
finitely many vertices and edges.

The wave function $\psi_t$ provides us with a probability density
\begin{equation}\label{rhodef}
  \rho_t(q) = |\psi_t(q)|^2
\end{equation}
and a probability current vector
\begin{equation}\label{jdef}
  j_t(q) = \hbar \, \Im \bigl( \psi_t^*(q)\, \nabla \psi_t(q) \bigr)\,,
\end{equation}
at least at every non-vertex $q \in \graph \setminus \vertices$. (In
case that the value space of $\psi_t$ is not $\CCC$ but a
higher-dimensional complex vector space $\CCC^n$, the product in the
bracket in \eqref{jdef} should be understood as an inner product in
$\CCC^n$.)  At a vertex $q$, there are several current vectors
\begin{equation}\label{jedindef}
  j_{\edin,t}(q) = \hbar \, \Im \bigl( \psi_t^*(q)\, \nabla_\edin
  \psi_t(q) \bigr)\,,
\end{equation}
one for each edge $\edin \in \edges_q$ ending there; this corresponds
to the fact that at $q$ there is one (one-sided) derivative
$\nabla_\edin$ for each edge $\edin \in \edges_q$.  Together, the
$j_t(q)$ for $q \in \graph \setminus \vertices$ and the
$j_{\edin,t}(q)$ for $q \in \vertices$ and $\edin \in \edges_q$ form
what can be regarded as a \emph{vector field}, denoted $j_t$, on
$\graph$, consisting of one element in each tangent space, where a
vertex is thought of as having several tangent spaces, one for each edge.

The obvious choice for the law of motion along an edge (outside the
vertices) is Bohm's \cite{Bohm52,DGZ92}, i.e., the deterministic law
\begin{equation}\label{Bohm}
  \frac{dQ_t}{dt} = v_t(Q_t) = \frac{j_t(Q_t)}{\rho_t(Q_t)} \,.
\end{equation}
To have $Q_t$ follow the vector field $v_t = j_t/\rho_t$ ensures that
the probability current $\rho_t v_t$ of the process (at non-vertices)
agrees with the prescribed current $j_t$, provided the process has
distribution $\rho_t$ as intended.

According to the probability distribution \eqref{rhodef}, the
probability of $Q_t \in \vertices$ vanishes, like for every other finite
subset of $\graph$; this suggests that whenever $Q_t$ reaches a vertex
$q$ it should leave $q$ immediately, rather than spend some time
sitting on $q$. Since we want that the probability flux of the process
be given by $j_t$ (and that the paths are continuous), we need that
the flux into the vertex is as large as the flux out of the vertex,
that is, that the net flux into the vertex is zero. This
\emph{Kirchhoff condition} can be expressed by the formula
\begin{equation}\label{sumfluxzero}
  \sum_{\edin \in \edges_q} n_\edin(q) \cdot j_{\edin,t}(q) = 0\,,
\end{equation}
where $n_\edin(q)$ is the unit vector at $q$ pointing in the direction
of the edge $\edin$ (that is, away from $q$), and the dot $\cdot$
denotes the inner product in the tangent space to the edge $\edin$
(with $\edin$ regarded as a Riemannian manifold with boundaries) at
the point $q$.  At a vertex at which just one single edge ends,
\eqref{sumfluxzero} requires the current to vanish.  The meaning of
the Kirchhoff condition \eqref{sumfluxzero} is local conservation of
probability at the vertex $q$; no probability gets lost or added.
Therefore it is the analogue, at the vertices, of the continuity
equation
\begin{equation}\label{continuity}
  \frac{\partial \rho_t}{\partial t} (q) = - \nabla \cdot j_t(q)\,,
\end{equation}
which expresses the local conservation law at non-vertices $q \in
\graph \setminus \vertices$. 

While \eqref{continuity}, with \eqref{rhodef} and \eqref{jdef}
inserted, is a consequence of the Schr\"odinger equation \eqref{schr},
\eqref{sumfluxzero} is an additional requirement.  The simplest way,
and presumably the only practical way, of ensuring \eqref{sumfluxzero}
for all times is to impose a boundary condition on the wave function
$\psi$ that implies \eqref{sumfluxzero}.\footnote{The term ``boundary
  condition'' is in a sense inappropriate, a sense in which ``vertex
  condition'' would be more appropriate: the condition concerns the
  behavior of $\psi$ at vertices, and vertices are not (necessarily)
  boundaries of the graph. They are boundaries, however, of the edges
  glued to them, and that is how the similarity with boundary
  conditions in other quantum mechanical situations comes about.}
Whereas \eqref{sumfluxzero} is a condition quadratic in $\psi$, the
boundary condition on $\psi$ should be linear: otherwise the
acceptable wave functions would not form a linear space, and there
would be little hope that the boundary condition could be conserved by
the evolution of the wave function. A natural choice of boundary
condition is thus a \emph{Robin boundary condition},
\begin{equation}\label{boundarycond}
  \alpha (q) \sum_{\edin \in \edges_q} n_\edin(q) \cdot \nabla_\edin
  \psi(q) = \beta(q) \, \psi(q) \,, \quad q \in \vertices\,,
\end{equation}
where $\alpha(q)$ and $\beta(q)$ are real constants (and not both zero). 
Here, it is assumed that
\begin{equation}\label{conticond}
  \psi \text{ is continuous at vertices},
\end{equation}
so that $\psi(q) \to \psi(q_0)$ as $q \to q_0 \in \vertices$ along an
edge. Together, \eqref{boundarycond} and \eqref{conticond} imply the
Kirchhoff condition \eqref{sumfluxzero} on the current just as the
Schr\"odinger equation \eqref{schr} implies the continuity equation
\eqref{continuity}. This is seen, if $\alpha(q) \neq 0$, by
multiplying \eqref{boundarycond} by $\alpha(q)^{-1} \, \hbar \,
\psi^*(q)$ and taking the imaginary part, observing that $\alpha(q)$
and $\beta(q)$ are real. In the case $\alpha(q) =0$,
\eqref{boundarycond} reduces to the \emph{Dirichlet boundary
  condition} $\psi(q) =0$, which also obviously implies the Kirchhoff
condition \eqref{sumfluxzero} on the current. (The Dirichlet condition
is the simplest condition at an external vertex (i.e., one with only
one edge) though not very interesting for us at internal vertices as
it excludes any flux of probability across the vertex.)

In fact, the Laplacian on the functions $\psi$ (with the right degree
of regularity, namely from the second Sobolev space on each edge) 
satisfying \eqref{boundarycond} and \eqref{conticond} is self-adjoint
\cite{vB1,vB2,KS99,kuchment}. Therefore, equations \eqref{schr},
\eqref{boundarycond}, and \eqref{conticond} together define a unitary
evolution on Hilbert space. The constants $\alpha(q)$ and $\beta(q)$
determine how much of an incoming wave gets reflected and how much
transmitted, and with what phase shift.

(A remark, in brackets, on the other self-adjoint extensions of the
Laplacian: On complex-valued functions, the extensions defined by \eqref{boundarycond} are, in fact, all 
self-adjoint extensions for continuous functions, 
i.e., assuming \eqref{conticond} \cite{kuchment,KS99}. Further 
self-adjoint extensions exist if one drops \eqref{conticond}; the most 
general local vertex condition defining a self-adjoint extension for 
complex-valued wave functions is the following \cite{kuchment,KS99}: 
(i)~Along each edge $\edin \in \edges_q$, a limit of $\psi$ in the vertex 
$q$ exists, $\psi(q,\edin) := \lim \psi(q')$ as $q' 
\to q$ along $\edin$, though the limits may differ for different edges.
(ii)~Let $F \in \CCC^d$, where $d= \# \edges_q$ is the
degree of the vertex $q$, be the vector with the components $F_\edin =
\psi(q,\edin)$, and $F' \in \CCC^d$ the vector
with the components $F'_\edin = n_\edin(q) \cdot \nabla_\edin \psi(q)$.
The further conditions are $P^\perp F =0$ and $PF' +LPF = 0$, where 
$P$ is an orthogonal projection in $\CCC^d$, $P^\perp = 1 - P$ the complementary
projection, and $L$ a self-adjoint endomorphism on the range of $P$. This
vertex condition includes \eqref{boundarycond} for $P$ the projection on
$(1,1,\ldots,1)$ and $L = \alpha(q)^{-1} \beta(q)$ (in case $\alpha(q) \neq 0$)
respectively $P=0$ (in case $\alpha(q) = 0$). This vertex condition also 
implies the Kirchhoff condition \eqref{sumfluxzero} on the flux, since the
left hand side of \eqref{sumfluxzero} equals, up to a factor $\hbar$, 
$\Im \langle F,F'\rangle = \Im \langle PF,F'\rangle = \Im \langle PF,
PF'\rangle = -\Im \langle PF,LPF\rangle = 0$, where $\langle \cdot,
\cdot \rangle$ denotes the scalar product in $\CCC^d$. The Kirchhoff
condition on the flux is all we need to be able to
define the minimal graph process. It is only for the sake of simplicity that
we restrict our attention to the simpler condition \eqref{boundarycond}.)

Now that we have made precise the evolution of the wave function
$\psi_t$, let us turn to the process $Q_t$.  Once $Q_t$ has reached a
vertex $q$ by its deterministic motion, a decision needs to be made
about along which edge to leave the vertex. Those edges $\edin$ for
which $n_\edin(q) \cdot j_{\edin,t}(q) < 0$ holds, do not possess
trajectories---solutions of \eqref{Bohm}---that begin at $q$ at time
$t$. Those edges $\edin$, in contrast, for which $n_\edin(q) \cdot
j_{\edin,t}(q) > 0$ holds, do.  The borderline case $n_\edin(q) \cdot
j_{\edin,t}(q) = 0$ we can ignore.  Thus, conditional on $Q_t = q \in
\vertices$, the simplest way of choosing the edge $\edin \in \edges_q$
along which to leave $q$ is to choose it at random with probability
\begin{equation}\label{edgeprob}
  \PPP_t(\edin|q) = \frac{[n_\edin(q) \cdot j_{\edin,t}(q)]^+}
  {\sum\limits_{\aedin \in \edges_q} [n_{\aedin}(q)
  \cdot j_{\aedin,t}(q)]^+} \,,
\end{equation}
where $x^+ = \max(x,0)$ denotes the positive part of $x \in \RRR$.
Note that by construction $\PPP_t(\edin|q) \geq 0$ and $\sum_{\edin
  \in \edges_q} \PPP_t(\edin|q) =1$. $\PPP_t(\edin|q)$ is ill-defined
when and only when the denominator vanishes (assuming that $j_t$ is
well defined), which happens, by \eqref{sumfluxzero}, when and only
when $j_{\edin,t}(q) =0$ for all $\edin \in \edges_q$.

This completes the definition of the minimal graph process: the wave
function $\psi_t$ evolves according to the PDE \eqref{schr} with the
boundary conditions \eqref{boundarycond} and \eqref{conticond}, and
the process $Q_t$ moves according to the ODE \eqref{Bohm} with the
stochastic law \eqref{edgeprob} at every vertex. We have left out of
consideration the possibility of a topological phase factor associated
with every non-contractible closed path (see \cite{topid1A} for a
discussion), as this possibility does not affect those features of the
process that we are interested in.

\section{Equivariant Processes}

The stochastic law \eqref{edgeprob}, and thus the minimal graph
process, is uniquely determined by the following requirements:
\begin{itemize}
\item[(i)] $(Q_t)_{t\in\RRR}$ is a Markov process, 
\item[(ii)] it has continuous paths, 
\item[(iii)] it is equivariant, i.e., $|\psi_t|^2$ distributed at
   every $t$, 
\item[(iv)] the motion along the edges is Bohmian, i.e., given by
   \eqref{Bohm}. 
\end{itemize}

To see this, consider a process satisfying (ii) and (iv). Such a
process, whenever it is in a vertex $q$, waits a random time in
$q$, which must almost surely be zero if it is equivariant, and then
selects at random one of the edges for leaving $q$. If it is a
Markov process, the probabilities for the various edges $\edin \in
\edges_{q}$ do not depend on the past history of the process, not
even on the edge along which it reached $q$; thus, they are given by a
function $\PPP_t (\edin|q)$ of time $t$ and vertex $q$ alone. The
probability flux (per time) out of $q$ along $\edin$ is then
\begin{equation}\label{outflux}
  J^\mathrm{out}_{\edin,t}(q) = \PPP_t (\edin|q)\,
  \sum_{\aedin\in\edges_q} J^\mathrm{in}_{\aedin,t}(q) \, ,
\end{equation}
where $J^\mathrm{in}_{\aedin,t}(q)$ is the probability flux (per time)
into $q$ along $\aedin$.

Now invoke the $|\psi|^2$ distribution.  The flux into $q \in
\vertices$ along $\edin \in \edges_{q}$ between $t$ and $t+dt$ equals
the $|\psi|^2$ measure of the Bohmian trajectories ending at $q$
between $t$ and $t+dt$.\footnote{\label{arrivalstatistics}That means,
  in case all the trajectories along $\edin$ ending at $q$ during
  $[t_1,t_2]$ existed already at time $t_0 \leq t_1$ (rather than
  started, at one of the ends of $\edin$, after $t_0$), that the flux
  into $q$ along $\edin$ during $[t_1,t_2]$ is
  \[
    \int_{t_1}^{t_2} J^\mathrm{in}_{\edin,t}(q) \, dt = \int_{\edin}
    |\psi_{t_0}(q')|^2 \, 1_{\{t_1 \leq \tau(q') \leq t_2\}}\, dq' 
  \]
  with $\tau(q')$ the time of arrival at $q$ (i.e., $\infty$ in case
  of no arrival) of the trajectory starting in $q'\in\edin$ at time
  $t_0$.}  It follows, since the arrival statistics of the Bohmian
trajectories at $q$ is given by the quantum current $j_{\edin,t}(q)$
defined in \eqref{jedindef}, that
\begin{equation}\label{equiinflux}
  J^\mathrm{in}_{\edin,t}(q) = j^\mathrm{in}_{\edin,t}(q) :=
  [n_\edin(q) \cdot j_{\edin,t}(q)]^- \,,
\end{equation}
where $x^- = \max(-x,0)$ denotes the negative part of $x \in \RRR$.
Similarly, to obtain the $|\psi|^2$ distribution along the edge
$\edin$ at later times, it is necessary that the amount of probability
leaving $q$ per unit time along $\edin$ is
\begin{equation}\label{equioutflux}
  J^\mathrm{out}_{\edin,t}(q) = j^\mathrm{out}_{\edin,t}(q) :=
  [n_\edin(q) \cdot j_{\edin,t}(q)]^+ \,.
\end{equation}
That is because no other trajectories can contribute to the
probability contents of any interval of $\edin$ than those which
started in the right time interval.  Inserting \eqref{equiinflux} and
\eqref{equioutflux} into \eqref{outflux}, and observing that by the
Kirchhoff condition \eqref{sumfluxzero} we have that
\begin{equation}\label{influxoutflux}
  \sum_{\edin \in \edges_q} j^\mathrm{in}_{\edin,t}(q) =
  \sum_{\edin \in \edges_q} j^\mathrm{out}_{\edin,t}(q) \,,
\end{equation}
we obtain \eqref{edgeprob}.

Note that while we required only equivariance, we obtained more,
namely the ``standard current property'' \cite{crea2A,crea2B,schwer}:
not only the distribution density of the process $Q_t$ agrees with the
quantum value \eqref{rhodef}, but also its probability current agrees
with the quantum current as given by \eqref{jdef} and
\eqref{jedindef}.

It is a corollary of this uniqueness result that generically, a
process satisfying our requirements (i)--(iv) \emph{cannot be
  deterministic}. (The exception is when the minimal graph process is
deterministic, which is when it so happens that at every vertex at
every time the outflux either vanishes or takes place along a single
edge.)  

In fact, already (ii) and (iii) alone are generically incompatible
with determinism. To see this, let $(Q_t)_{t\in\RRR}$ be a
deterministic process with continuous paths on the graph consisting of
three copies $\edin_1, \edin_2, \edin_3$ of $[0,\infty)$ joined at a
single vertex $q$. We show that $Q_t$ cannot be equivariant for wave
functions $\psi$ such that, during a time interval $[t_1,t_2]$,
$n_{\edin_1}(q) \cdot j_{\edin_1,t}(q) <0$ and $n_{\edin_i}(q) \cdot
j_{\edin_i,t}(q) >0$ for $i=2,3$. The reason is essentially that, due
to determinism, the flux into $q$ along $\edin_1$ leaves $q$, at every
time, along either $\edin_2$ or $\edin_3$ but not both, though it
would have to be split to maintain equivariance.

In detail, with the notation $P_i (t)= \int_{\edin_i} |\psi_t(q')|^2
\, dq'$ for the $|\psi|^2$ measure of $\edin_i$, we have that $P_1$ is
strictly decreasing since $(d/dt)P_1 = n_{\edin_1}(q) \cdot
j_{\edin_1,t}(q) <0$ whereas $P_2$ and $P_3$ are strictly increasing,
$(d/dt) P_i > 0$, $i=2,3$.  By continuity, a path $t\mapsto Q_t$ can
pass from one edge to another only by crossing $q$. Let $\edin(t)$ and
$\aedin(t)$ be the edges along which the path crossing $q$ at time $t$
reaches respectively leaves the vertex; this is well defined due to
the assumed determinism.  Let $J_{i}(t) \, dt$ be the probability that
the path $s\mapsto Q_s$ crosses the vertex between $t$ and $t+dt$ and
leaves along $\edin_i$. With the notation $R_i(t) := \mathrm{Prob}(Q_t
\in \edin_i)$ for the probability contents of $\edin_i$, we have that
for every subinterval $[t_3,t_4] \subseteq [t_1,t_2]$,
\begin{equation}\label{fluxtoedge}
  R_i(t_4) \leq R_i(t_3) + \int_{t_3}^{t_4} J_{i}(t) \, dt \,.
\end{equation}
Let $S_{i}$ be the set of $t \in [t_1,t_2]$ for which $\aedin(t) =
\edin_i$.  Since during $S_{2}$ no paths can enter $\edin_3$,
$\int_{S_{2}} J_{3}(t) \, dt = 0$.  If $Q_t$ were equivariant, then
$R_i(t) = P_i(t)$ for all $t \in [t_1,t_2]$ and, by
\eqref{fluxtoedge}, $(d/dt) P_i \leq J_{i}$, and thus $J_2(t)>0$ and
$J_3(t) > 0$. Therefore, $S_{2}$ would have to be a null set, and thus
$R_2(t_2) \leq R_2(t_1) + \int_{S_2} J_2(t) \, dt = R_2(t_1) =
P_2(t_1) < P_2(t_2)$, in contradiction to equivariance.

Let us turn again to indeterministic processes. Other processes than
the minimal graph process are possible when we drop the Markov
property from our requirements.  Then the distribution of the outgoing
edge can depend on the past history of the process, and the most
interesting possibility is perhaps that it is a function $\PPP_t
(\edin|q,\aedin)$ of the edge $\aedin$ along which $q$ was reached,
yielding what could be called an almost-Markovian process. The
condition on $\PPP_t (\edin|q,\aedin)$ deriving from (ii)--(iv) is
\begin{equation}\label{ededprob}
  \sum_{\aedin \in \edges_q} \bigl[ n_\aedin(q) \cdot j_{\aedin,t}(q)
  \bigr]^- \, \PPP_t (\edin|q,\aedin) = \bigl[ n_\edin(q)
  \cdot j_{\edin,t}(q) \bigr]^+ \,,
\end{equation}
together with
\begin{equation}\label{edednorm}
  \sum_{\edin \in \edges_q} \PPP_t (\edin|q,\aedin) =1\,,
\end{equation}
and this is an underdetermined system of equations for the quantities
$\PPP_t (\edin|q,\aedin)$ whenever there is influx along more than one
edge and outflux along more than one edge.

Further equivariant processes are possible if we drop the requirement
(iv) that the motion along the edges is Bohmian. One could consider
instead an equivariant diffusion process such as Nelson's
\emph{stochastic mechanics} \cite{stochmech2,stochmech1}. Diffusion
processes can be defined on a graph as well \cite{vB1,vB2}; in
order to specify such a process, one has to specify the diffusion
constant and drift for every $q\in\graph$ and time $t$, and, in
addition, for every vertex $q$ and time $t$ the probability
distribution $\PPP_t(\edin|q)$ on $\edges_q$ for which edge to select
upon arrival at $q$.\footnote{The situation concerning the outgoing
  edge is more subtle though, as a diffusion process in one dimension
  returns to its starting point infinitely often within arbitrarily
  short times, so that one cannot speak of the ``next edge'' that the
  process enters after being in a vertex.}  With stochastic mechanics
along the edges, the biggest difference for the uniqueness question is
that the process can leave $q$ along $\edin$ even if $n_\edin (q)
\cdot j_{\edin,t}(q) < 0$, and this opens up a lot of freedom.

\section{Comparison with Bell-Type QFT}

We now contrast the minimal graph process with Bell-type quantum field
theories: these involve Markov processes on spaces $\Q$ that are
countable unions of disjoint manifolds (typically representing the
configuration space of a variable number of particles), with
stochastic jumps and deterministic, continuous (Bohmian) trajectories
in between.  The jumps, discontinuities in the path, occur at random
times and lead to random destinations, with jump rates given by a
formula in terms of $\psi$ analogous to Bell's \cite{crea2A}. Upon
arrival at the destination, there is only one possibility, unlike at a
graph vertex, in which direction the process can move on: it is
defined by the Bohmian velocity vector field on $\Q$ at the
destination. Outside the context of quantum theory, similar Markov 
processes on the configuration space of a 
variable number of particles, representing many interacting particles
with fragmentation and coagulation at random times, have been
considered in statistical mechanics \cite{BK03} and probability theory 
\cite{preston}.

The stochastic jumps in a Bell-type quantum field theory correspond to
the term $H_I$ in the Hamiltonian $H = H_0 + H_I$, where $H_0$, a
differential operator, is called the ``free Hamiltonian'' and $H_I$,
an integral operator, the ``interaction Hamiltonian''. In fact, this splitting of the 
Hamiltonian corresponds to a splitting $L =L_0 + L_I$ of the generator 
$L$ of the Markov process, where $L_0$ generates the continuous motion
and $L_I$ the jumps \cite{crea2B} (for a general discussion of such a splitting 
of Markov generators see \cite{BK03}), so that one could
consider a process for $H_0$ alone, which will be a process without
jumps (in fact just the Bohmian motion). For the minimal graph process, in contrast,
there does not exist a comparable splitting of the Hamiltonian in two
contributions such that one would correspond to deterministic motion
and the other to the stochastic decisions.  There is, however, a
correspondence we can make: the Hamiltonian is defined by a
differential operator and a boundary condition; the differential
operator corresponds to the deterministic Bohmian motion, while the
boundary condition corresponds to the stochastic decision made at
every vertex.

\section{A Limiting Case of Bell's Process}

It is well known \cite{Sudbery,Vink,Colin} (though not on a rigorous
level) that Bohmian mechanics is a limiting case of Bell's process:
you approximate Euclidean space $\RRR^n$ by a lattice $\varepsilon \ZZZ^n$
and the Laplacian by the lattice Laplacian, consider Bell's process
and let $\varepsilon \to 0$. We now derive the minimal graph process
as the limiting process of a suitable approximation by means of Bell's
process. To this end, we replace each edge isometric to $[0,a]$ by a
lattice $[0,a] \cap \varepsilon \ZZZ$ and the Laplacian by the lattice
Laplacian.  Since we already know that along the edge, Bell's process
converges to Bohmian mechanics, what remains to be investigated is the
behavior of Bell's process at the vertices (which we include among the
lattice sites). The rate for jumping from the vertex $q$ to the
nearest site along the edge $\edin$, which we denote symbolically by
$q+ \varepsilon \edin$, is \cite{Bell86,crex1,crea2B}
\begin{equation}
  \sigma_t(q+\varepsilon \edin|q) = \frac{\bigl[\tfrac{2}{\hbar}
  \Im \, \psi_t^*(q+\varepsilon \edin) \,
  \sp{q+ \varepsilon \edin}{H|q}
  \, \psi_t(q) \bigr]^+}{\psi^*_t(q) \, \psi_t (q)} \,.
\end{equation}
Since $\sp{q+ \varepsilon \edin}{H|q} = \hbar^2/2\varepsilon^2$ for
the lattice Laplacian, and since $\psi(q + \varepsilon \edin) -
\psi(q)$ is of the order $\varepsilon$, the jump rate is of the order
$\varepsilon^{-1}$; thus, the waiting time is of the order
$\varepsilon$ and goes to zero in the continuum limit $\varepsilon \to
0$. We are interested in the distribution over the edges. The
probability that the process leaves $q$ along $\edin$ is
\[
  \frac{\sigma_t(q+\varepsilon \edin|q)}{\sum\limits_{\aedin \in \edges_q}
  \sigma_t(q+\varepsilon \aedin|q)} = 
  \frac{\bigl[\tfrac{2}{\hbar}
  \Im \, \psi_t^*(q+\varepsilon \edin) \, \sp{q+ \varepsilon \edin}{H|q}
  \, \psi_t(q) \bigr]^+}{\sum\limits_{\aedin \in \edges_q}
  \bigl[\tfrac{2}{\hbar}
  \Im \, \psi_t^*(q+\varepsilon \aedin) \, \sp{q+ \varepsilon \aedin}{H|q}
  \, \psi_t(q) \bigr]^+} =  \frac{[n_\edin(q) \cdot j_{\edin,t}(q)]^+}
  {\sum\limits_{\aedin \in \edges_q} [n_{\aedin}(q)
  \cdot j_{\aedin,t}(q)]^+} \,,
\]
where $t$ is the time at which the process leaves $q$, and $n_\edin(q)
\cdot j_{\edin,t}(q)$ denotes the probability flux out of $q$ into
$\edin$, which in the lattice model equals $\tfrac{2}{\hbar} \Im \,
\psi_t^*(q+\varepsilon \edin) \, \sp{q+ \varepsilon \edin}{H|q} \,
\psi_t(q)$. This obviously converges to \eqref{edgeprob}, as we claimed.

\section{Sort of Limiting Case of Bohmian Motion}

Consider a graph $\graph$ isometrically embedded in $\RRR^n$ and
Bohmian mechanics in $\RRR^n$ with a potential $V$ that forces the
particle to stay $\varepsilon$-close to $\graph$. As we take the limit
$\varepsilon \to 0$, does the process converge to the minimal graph
process? The answer is in general no, but it seems plausible that the
Markovization of the limiting process is the minimal graph process.
The following example illustrates what happens.

Take $n=2$ and $\graph$ consisting of the four half-axes with the
origin as the only vertex. An example potential that keeps $Q_t$ close
to $\graph$ is $V(x,y) = \min \{(x/\varepsilon)^2,
  (y/\varepsilon)^2\}$. Every point $q$ in the plane minus the two
diagonals possesses a unique closest point $\pi(q)$ on $\graph$:
indeed, taking out the two diagonals decomposes the plane into four
quadrants, each containing one half-axis, and $\pi$ on a quadrant is
the orthogonal projection to that half-axis.  If $Q_t^\varepsilon$ is
the Bohmian path, $\pi (Q_t ^\varepsilon)$ is a process on $\graph$,
and one could imagine that for suitable choice of the initial wave
function $\psi_0 ^\varepsilon: \RRR^2 \to \CCC$ as a function of
$\varepsilon$, $\pi (Q_t ^\varepsilon)$ possesses a limiting process
$Q_t^0$ as $\varepsilon \to 0$. One could also imagine that the
discontinuity that occurs in $t \mapsto \pi (Q_t ^\varepsilon)$
whenever $Q_t ^\varepsilon$ crosses a diagonal vanishes in the
limit, as $Q_t ^\varepsilon$ crosses the diagonal in an
$\varepsilon$-neighborhood of the origin.

However, the transversal coordinate of $Q_t^\varepsilon$, the one that
is projected out by $\pi$, may decide about the edge along which to
leave the central region, as depicted in Figure~1. As a consequence,
the probability distribution for the outgoing edge may depend on the
ingoing edge, so that the limiting process $Q_t^0$ is not Markovian,
but instead the kind of almost-Markovian process described by
\eqref{ededprob} and \eqref{edednorm}.

\begin{figure}[ht]
\begin{center}
\includegraphics[width=.5 \textwidth]{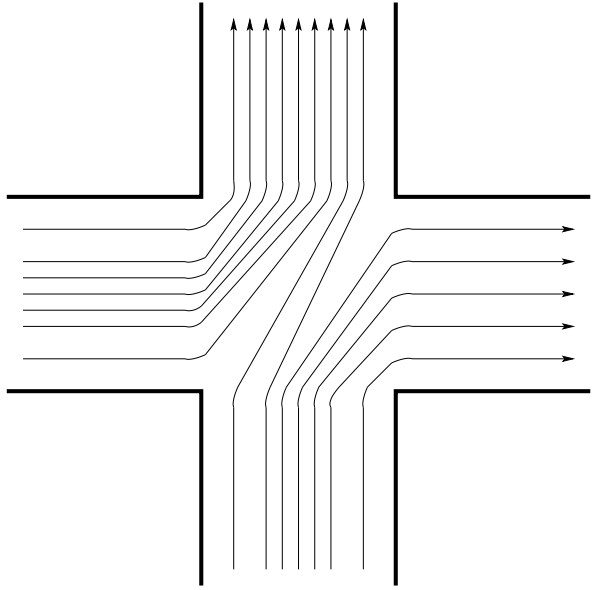}
\end{center}
\caption{The Bohmian trajectories, qualitatively, in an example case
  in which the motion is confined to an $\varepsilon$-neighborhood of
  the axes (bounded by the bold lines). In the limit $\varepsilon \to
  0$, the motion takes place along the axes, and may turn from one
  axis to the other at the origin.  In this example, trajectories
  coming in from the left go out upwards whereas trajectories coming
  in from below go out either upwards or to the right; therefore, the
  projection $\pi(Q_t)$ to the axes is not Markovian.}
\label{figone}
\end{figure}

Still, a process of this kind has the property that its Markovization
is the minimal graph process (for the same wave function). The
\emph{Markovization} of a stochastic process $Q_t$ is defined as the
Markov process $\tilde{Q}_t$ with
\begin{equation}\label{Markovization}
  \prob \bigl( \tilde{Q}_t \in B \text{ and } \tilde{Q}_{t+dt} \in C \bigr)
  = \prob \bigl( Q_t \in B \text{ and } Q_{t+dt} \in C \bigr)\,,
\end{equation}
for all sets $B,C$ and all $t$.  It is not obvious that a
Markovization exists. In contrast, in discrete time the Markovization,
with $dt$ replaced by the time step, obviously exists and is unique in
law. When $\tilde{Q}_t$ exists, it has the same one-time marginals and
the same transition probabilities (not conditional on the prior
history) for infinitesimal time differences $dt$.

For an almost-Markovian process moving with velocities $v_t$ along the
edges and selecting the outgoing edge $\edin$ at a vertex $q$ reached
along $\aedin$ with distribution $\PPP_t(\edin|q,\aedin)$, the
Markovization looks as follows: it moves again with velocities $v_t$
along the edges and selects the outgoing edge at a vertex $q$ with
distribution $\tilde{\PPP}_t(\edin|q)$ given by
\begin{equation}
  \tilde{\PPP}_t (\edin|q) = \frac{ \sum\limits_{\aedin \in \edges_q}
  \PPP_t (\edin|q,\aedin) \, \rho_{\aedin,t}(q)\, [n_{\aedin}(q)
  \cdot v_{\aedin,t}(q)]^-} {\sum\limits_{\aedin \in \edges_q}
  \rho_{\aedin,t}(q)\, [n_{\aedin}(q) \cdot v_{\aedin,t}(q)]^- } \,.
\end{equation}
If the process is equivariant, $v_t$ is the Bohmian velocity vector
field, and $\PPP_t(\edin|q,\aedin)$ satisfies the equivariance
condition \eqref{ededprob}, then, by the Kirchhoff condition
\eqref{sumfluxzero}, the distribution of the edges $\tilde{\PPP}_t
(\edin|q)$ of the Markovization equals the one of the minimal graph
process, \eqref{edgeprob}.

\section{Symmetries}

The minimal graph process respects the symmetries of the Schr\"odinger
equation. That is, suppose that $\graph$ possesses an isometry
$\varphi: \graph \to \graph$. Then $\varphi(Q_t)$ is again a minimal
graph process, associated with the wave function $\psi \circ \varphi$,
which obeys the Schr\"odinger evolution with potential $V \circ
\varphi$. If $V$ is symmetric under $\varphi$, one obtains in this way 
further solutions $\psi_t,Q_t$ of the same set of defining equations \eqref{schr}, 
\eqref{Bohm}, and \eqref{edgeprob}. If, in addition, $\psi_t$ is symmetric 
under $\varphi$, it remains so for all times, and the distribution of the
process $(Q_t)_{t \in \RRR}$, regarded as a measure on the path space,
is symmetric under the action of $\varphi$. Note that the
isometries of $\graph$ always form a finite set (of at most $k!2^k$ elements 
if the graph has $k = \# \edges$ edges, since every isometry defines a
permutation of the edges, and there are only two ways of isometrically 
mapping one edge to another one of the same length).

Another symmetry the minimal graph process respects is time reversal:
$(Q_{-t})_{t \in \RRR}$ is again a minimal graph process, associated
with the wave function $\psi_t' = \psi_{-t}^*$. (As usual in quantum
mechanics, the wave function has to be replaced, under time reversal,
by its complex conjugate.  Then $\psi'$ solves the Schr\"odinger
equation again.) To see this, note that, as a consequence of the
conjugation, the currents \eqref{jdef} change sign while the density
$|\psi|^2$ remains unchanged. Therefore, the Bohmian velocities
\eqref{Bohm} change sign, as they should. If the process is in a
vertex $q$ at time $t$, then the probability that it came along
$\aedin$ and leaves along $\edin$ is, by \eqref{edgeprob},
proportional to $[n_\aedin(q) \cdot j_{\aedin,t}(q)]^- \, [n_\edin(q)
\cdot j_{\edin,t}(q)]^+$. Therefore, \eqref{edgeprob} holds again for
the reversed process with the reversed currents.

\section{Topology and Stochasticity}

We can regard graphs as test cases, or toy models, for the more
complicated spaces of higher dimension arising from gluing together
parts of Euclidean spaces or manifolds. Such spaces are not at all
eccentric as configuration spaces; especially for configuration spaces
for a variable number of particles, is it a natural thought to
identify certain configurations, and thus to glue together different
parts of the configuration space, initially given as disjoint
manifolds. For example, one may think of identifying the configuration
$q$ consisting of an electron at $\boldsymbol{x} \in \RRR^3$ and a
photon at the same location $\boldsymbol{x}$ with the configuration
$q'$ consisting of just an electron at $\boldsymbol{x}$. In this way,
the act of absorption no longer corresponds to a discontinuity in
configuration space. As another, though similar, example, one may
identify the configuration $q$ consisting of a particle at
$\boldsymbol{x}$ and an anti-particle at the same location
$\boldsymbol{x}$ with the vacuum configuration. Such glued
configuration spaces have been considered in \cite{Bala1,Bala2} for a
study of the spin--statistics connection.

Taking graphs as test cases, a feature we observe is that the special
topological situation we encounter at vertices (shapes like
\textsf{Y}, $+$, $*$ etc.) inevitably leads to stochasticity.  This
could be connected to the stochasticity of Bell-type quantum field
theories, which is associated with the annihilation and even more with
the creation of particles.  After all, creation and annihilation
events involve crossing from one sector of configuration space to
another (corresponding to a different particle number), and if, as we
suggested above, the sectors are glued together, then one should
expect stochasticity exactly at the annihilation and creation events.

\bigskip

\noindent \textit{Acknowledgments.}  I thank Delio Mugnolo 
(Universit\`a di Bari, Italy), Rainer Nagel and Stefan Teufel 
(Eberhard-Karls-Universit\"at T\"ubingen, Germany), Joachim von
Below (Universit\'e du Littoral in Calais, France), and Lucattilio Tenuta
(SISSA in Trieste, Italy), for helpful
discussions. I am grateful for the hospitality of the Institut des
Hautes \'Etudes Scientifiques at Bures-sur-Yvette, France, where part
of the work on this paper was done.  This work has been partially
supported by the European Commission through its 6th Framework
Programme "Structuring the European Research Area" and the contract
Nr. RITA-CT-2004-505493 for the provision of Transnational Access
implemented as Specific Support Action.

\end{document}